\documentclass[a4paper]{spie}  
\usepackage[]{graphicx}

\title{Phase Characteristics of the ALMA 3 km Baseline Data}

\author{Satoki Matsushita\supit{a}, Yoshiharu Asaki\supit{b,c},
	Ryohei Kawabe\supit{d}, Ed Fomalont\supit{e,f},
	Denis Barkats\supit{f},\\
	and Stuartt Corder\supit{f}
\skiplinehalf
\supit{a}Academia Sinica Institute of Astronomy and Astrophysics,
	11F of Astronomy-Mathematics Building, AS/NTU, No.1, Sec. 4,
	Roosevelt Rd, Taipei 10617, Taiwan, R.O.C.; \\
\supit{b}Institute of Space and Astronautical Science,
	3-1-1 Yoshinodai, Chuou, Sagamihara, Kanagawa 252-5210, Japan; \\
\supit{c}Department of Space and Astronautical Science,
	The Graduate University for Advanced Studies, 3-1-1 Yoshinodai,
	Chuou, Sagamihara, Kanagawa 252-5210, Japan; \\
\supit{d}National Astronomical Observatory of Japan, 2-21-1 Osawa,
	Mitaka, Tokyo 181-8588, Japan; \\
\supit{e}National Radio Astronomy Observatory, Charlottesville,
	VA 22903, USA; \\
\supit{f}Joint ALMA Observatory, Alonso de C\'ordova 3107, Vitacura
	763 0355, Santiago, Chile
}

\authorinfo{Further author information: (Send correspondence to S.M.)\\
	S.M.: E-mail: satoki@asiaa.sinica.edu.tw, Telephone: 886 (0)2 3365 2200}

\begin{document} 
  \maketitle 

\begin{abstract}
  We present the phase characteristics study of the Atacama Large
  Millimeter/submillimeter Array (ALMA) long (up to 3~km) baseline,
  which is the longest baseline tested so far using ALMA.
  The data consist of long time-scale (10 -- 20 minutes) measurements
  on a strong point source (i.e., bright quasar) at various frequency
  bands (bands 3, 6, and 7, which correspond to the frequencies of
  about 88 GHz, 232 GHz, and 336 GHz).
  Water vapor radiometer (WVR) phase correction works well even at
  long baselines, and the efficiency is better at higher PWV
  ($>1$~mm) condition, consistent with the past studies.
  We calculate the spatial structure function of phase fluctuation,
  and display that the phase fluctuation (i.e., rms phase) increases
  as a function of baseline length, and some data sets show turn-over
  around several hundred meters to 1~km and being almost constant at
  longer baselines.
  This is the first millimeter/submillimeter structure function at
  this long baseline length, and to show the turn-over of the
  structure function.
  Furthermore, the observation of the turn-over indicates that even
  if the ALMA baseline length extends to the planned longest
  baseline of 15~km, fringes will be detected at a similar rms phase
  fluctuation as that at a few km baseline lengths.
  We also calculate the coherence time using the 3 km baseline data,
  and the results indicate that the coherence time for band 3 is
  longer than 400 seconds in most of the data (both in the raw and
  WVR-corrected data).
  For bands 6 and 7, WVR-corrected data have about twice longer
  coherence time, but it is better to use fast switching method to
  avoid the coherence loss.
\end{abstract}



\section{INTRODUCTION}
\label{sect-intro}  

Atacama Large Millimeter/submillimeter Array (ALMA)\cite{hil10} has
recently become the world's largest millimeter-/submillimeter-wave
(mm-/submm-wave) interferometer, and currently commissioning and
Early Science observations are ongoing in parallel.
One of the most important commissioning items before the full
science operation of ALMA is the commissioning of long baselines.

The difficulty of long baseline interferometry in mm-/submm-wave
comes mainly from the large phase fluctuation caused by water vapor
in the atmosphere.
Refraction of water vapor causes the change of phase (and amplitude)
of electromagnetic wave.
Water vapor is believed to be distributed as clumps that have a
power-law size distribution and an upper limit of size at some point.
The difference of path lengths through water vapor clumps in the
line-of-sight between two antennas causes a phase difference, and the
change of this difference caused by the flow of water vapor clumps
creates phase fluctuation.
Due to the power-law size distribution and the upper limit of the
size of the water vapor clumps, the phase fluctuation is expected to
increase with baseline length, and at some point, phase fluctuation
does not become larger any more (see Ref.~\citenum{tho01} for more
details).

This phase fluctuation can be reduced using the 183~GHz Water Vapor
Radiometer (WVR)\cite{nik13}.
WVRs are installed in each of the 12-m diameter ALMA antennas, and
they measure the line-of-sight atmospheric water vapor content
precisely.
The difference of the water vapor content between antennas
corresponds to the difference of path lengths due to the water vapor
clumps, and therefore it is possible to estimate the phase difference
between antennas.
Correcting this estimated phase difference will correct the phase
fluctuation caused by water vapor in the atmosphere.
This phase correction method using WVRs has been shown to work
successfully on the past shorter baseline ALMA
data\cite{mat12,nik13}.

For the commissioning of ALMA long baseline, it is important to know
how large the phase fluctuation can be as a function of baseline
length, at which baseline length the phase fluctuation turns over
(i.e., stops increasing), and whether the WVR phase correction works
well even at the long baseline.
In this paper, we discuss these topics.

\section{MEASUREMENTS AND DATA ANALYSIS}
\label{sect-meas}  

\begin{figure}
\begin{center}
\begin{tabular}{c}
\includegraphics[width=16cm]{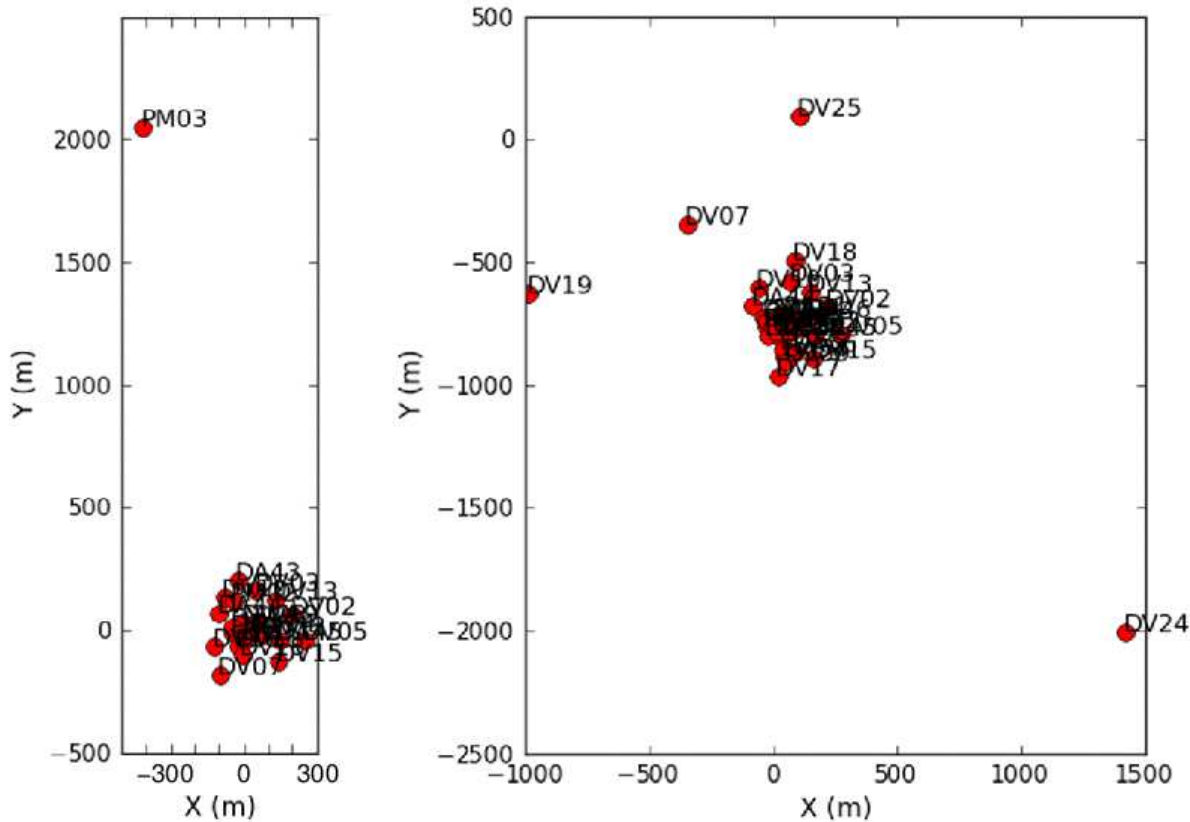}
\end{tabular}
\end{center}
\caption{
	\label{fig-antconf}
	{\it Left:} Antenna configuration for the 2~km baseline test.
	{\it Right:} Antenna configuration for the 3~km baseline test.
	}
\end{figure} 

To characterize the mm-/submm-wave phase fluctuation at long
baselines (up to 3~km), we stared at a strong point source, namely a
strong radio-loud quasar, using ALMA for $10-20$~minutes at frequency
bands 3 ($\sim88$~GHz), 6 ($\sim232$~GHz), and 7 ($\sim336$~GHz)
under various precipitable water vapor (PWV) conditions of
$0.5-2.0$~mm.
Note that the PWV values are from the WVR output, so that those are
the values along the line-of-sight.
We only used the 12~m diameter antennas, since the 7~m antennas do
not have WVRs.

There were two long baseline campaigns, one was for the maximum
baseline of $\sim3$~km performed on June 2013, and another of
$\sim2$~km on May 2012.
For both campaigns, most of the antennas were located at the central
cluster with the longest baselines of $\sim500$~m, but for the former
case, three antennas were located at north (antenna name DV25),
north-west (DV07), and west (DV19) of the central cluster with the
distance of about 1~km from the central cluster, and one antenna
located at south-east (DV24) with the distance of about 2~km (see
Fig.~\ref{fig-antconf} Right for the actual antenna configuration).
For the latter case, one antenna (PM03) was located at north with
the distance of about 2~km from the central cluster
(Fig.~\ref{fig-antconf} Left).
The details of the measurements are listed in Table~\ref{tab-meas3km}
and \ref{tab-meas2km} for the 3~km and 2~km baseline tests,
respectively.

\begin{table}[t]
\caption{Details of the 3~km baseline measurements.
	(1) Data identification number.
	(2) ALMA band name (number).
	(3) Measurement frequency in GHz.
	(4) Measurement date.
	(5) Precipitable water vapor in mm.
	(6) Observed source name.
	(7) Total observation time in minute.
	(8) Integration time of each data point in second.
	(9) Number of antennas used in the measurement.
	}
\label{tab-meas3km}
\begin{center}
\begin{tabular}{|c|cc|cccccc|}
\hline
\rule[-1ex]{0pt}{3.5ex}
 Data & Band & Freq. &     Date     & PWV  &  Source  & Obs.~Time & Int.~Time & Ant. \\
 No.  &      & [GHz] & [YYYY/MM/DD] & [mm] &   Name   &   [min]   &   [sec]   & No. \\
\hline
 1 & 3 &  88.0 &  2013/06/03  & 0.68 & 1058+015 &    9.6    &    0.96   & 28 \\
 2 & 3 &  88.0 &  2013/06/06  & 0.54 & NRAO 530 &    9.6    &    0.96   & 29 \\
 3 & 3 &  88.1 &  2013/06/06  & 0.55 & 1924-292 &    9.6    &    0.96   & 28 \\
 4 & 3 &  88.1 &  2013/06/06  & 0.55 & 2348-165 &    9.6    &    0.96   & 26 \\
 5 & 6 & 232.4 &  2013/06/06  & 0.54 & 2348-165 &    9.6    &    0.96   & 26 \\
 6 & 7 & 335.6 &  2013/06/06  & 0.57 & 2348-165 &    9.6    &    0.96   & 26 \\
 7 & 3 &  88.1 &  2013/06/07  & 1.46 & 1924-292 &    9.6    &    0.96   & 32 \\
 8 & 6 & 232.4 &  2013/06/07  & 1.42 & 1924-292 &    9.6    &    0.96   & 32 \\
 9 & 7 & 335.6 &  2013/06/07  & 1.39 & 1924-292 &    9.6    &    0.96   & 32 \\
\hline
\end{tabular}
\end{center}
\end{table} 

\begin{table}[t]
\caption{Details of the 2~km baseline measurements.}
\label{tab-meas2km}
\begin{center}
\begin{tabular}{|c|cc|cccccc|}
\hline
\rule[-1ex]{0pt}{3.5ex}
 Data & Band & Freq. &     Date     & PWV  &  Source  & Obs.~Time & Int.~Time & Ant. \\
 No.  &      & [GHz] & [YYYY/MM/DD] & [mm] &   Name   &   [min]   &   [sec]   & No. \\
\hline
 1 & 6 & 232.3 &  2012/05/02  & 1.54 & 1924-292 &   19.2    &    0.96   & 14 \\
 2 & 6 & 232.3 &  2012/05/02  & 1.65 & 1924-292 &   19.2    &    0.96   & 14 \\
 3 & 6 & 232.3 &  2012/05/10  & 0.98 &  3C279   &   19.2    &    0.96   & 15 \\
 4 & 6 & 232.3 &  2012/05/10  & 0.86 & 2258-279 &   19.2    &    0.96   & 14 \\
 5 & 3 &  88.0 &  2012/05/10  & 1.96 & 0522-364 &   19.2    &    0.96   & 17 \\
 6 & 3 &  88.0 &  2012/05/11  & 1.78 & 1924-292 &   19.2    &    0.96   & 15 \\
 7 & 3 &  88.0 &  2012/05/12  & 1.91 &  3C279   &   19.2    &    0.96   & 17 \\
 8 & 3 &  88.0 &  2012/05/12  & 1.14 &  3C454.3 &   19.2    &    0.96   & 14 \\
 9 & 7 & 335.6 &  2012/05/12  & 1.30 &  3C454.3 &   19.2    &    0.96   & 14 \\
10 & 3 &  88.0 &  2012/05/15  & 1.69 &  3C279   &   19.2    &    0.96   & 21 \\
11 & 3 &  95.8 &  2012/05/15  & 1.56 &  3C279   &   21.1    &    1.056  & 20 \\
12 & 3 &  95.8 &  2012/05/15  & 1.46 &  3C279   &   21.1    &    1.056  & 19 \\
13 & 3 &  95.8 &  2012/05/15  & 0.69 &  3C454.3 &   21.1    &    1.056  & 19 \\
14 & 3 &  88.0 &  2012/05/26  & 1.97 & 1924-292 &   19.2    &    0.96   & 16 \\
\hline
\end{tabular}
\end{center}
\end{table} 

All the data sets have been reduced under the Common Astronomy
Software Applications (CASA)\cite{mcm07} package environment.
The WVR phase correction has first been applied to the obtained data
sets using the {\tt wvrgcal} program\cite{nik13}.
For the 3 km baseline test data sets, the precise position for the
south-east antenna (DV24) was determined after these measurements,
so that we applied the correct antenna position at this point.

For the phase characterization, we calculate the spatial structure
function (hereafter SSF) of the phase fluctuation (i.e.,
root-mean-square [rms] phase) using personally developed python
programs under CASA.
The SSF of rms phase for each baseline can be calculated as
\begin{equation}
\sigma_\phi = \sqrt{\frac{<\{\phi(x+d) - \phi(x)\}^2>}{2}},
\end{equation}
where $\phi$ is a phase output from an antenna, $x$ is an arbitrary
antenna location, $d$ is the baseline length, and $<...>$ means an
ensemble average\cite{tat61,hol95,tho01}.
Original data output has an integration time of $\sim1$ second (see
Tables~\ref{tab-meas3km},\ref{tab-meas2km}), but for this
calculation, we averaged for 10 seconds to suppress noise with
white-spectrum characteristics such as that intrinsically arising
within the mixers of the ALMA receivers when measuring the phase from
the astronomical signals\cite{sra06}.
For the 10 second data binning, if the data points are less than 70\%
of the number of the data points should be, this binned data point
has been discarded.
Since there are limited number of baselines, especially for long
baselines, we did not bin the data along the baseline length.


\section{Effectiveness of Water Vapor Radiometer Phase Correction}
\label{sect-wvr}

\begin{figure}
\begin{center}
\begin{tabular}{c}
\includegraphics[width=16cm]{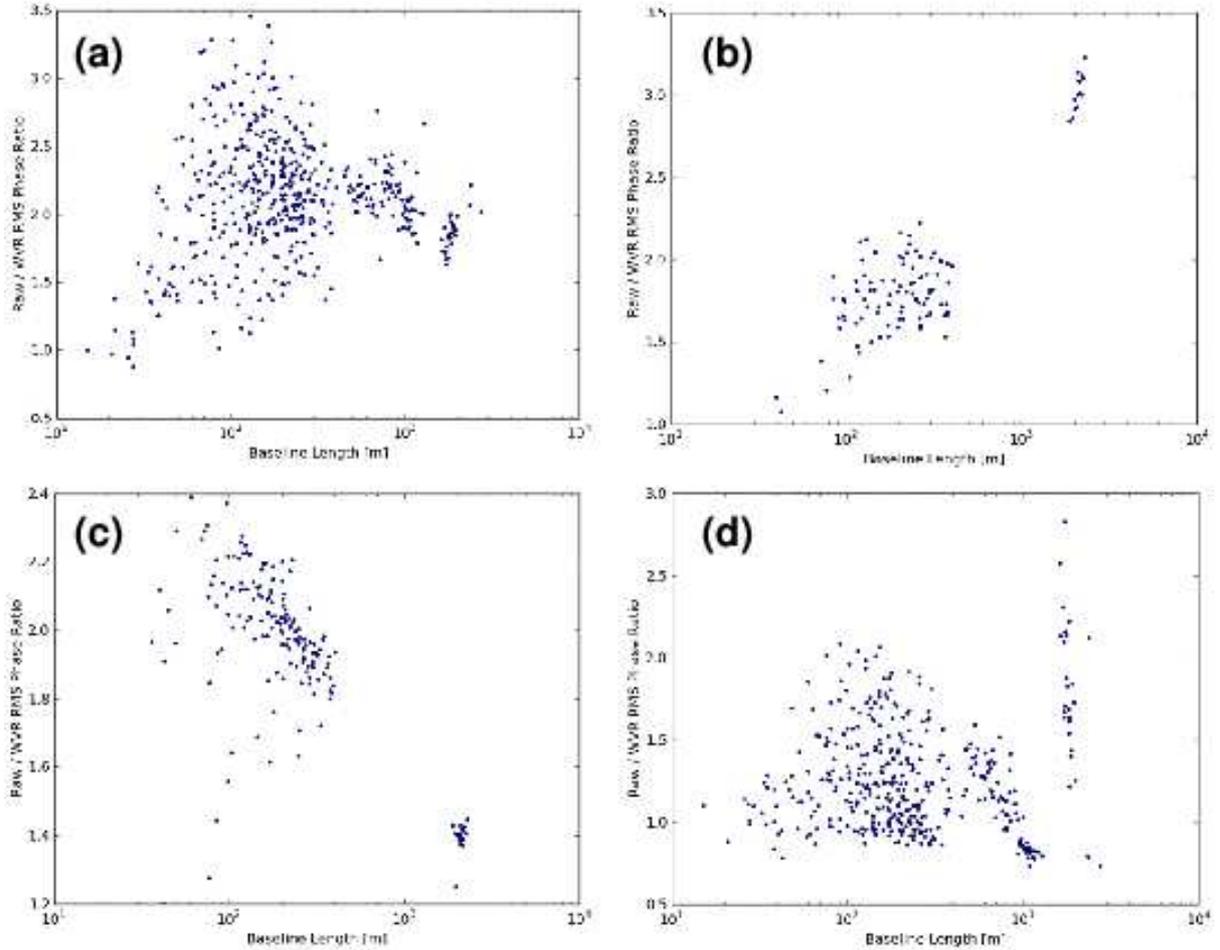}
\end{tabular}
\end{center}
\caption{
	\label{fig-ratio-bl}
	Example plots of the phase improvement ratio (the ratio of
	averaged rms phase between the raw and the WVR corrected data)
	as a function of baseline length.
	(a) A case of the ratio increases up to the baseline length of
		several hundred meters, and decreases at longer baselines.
		Most of the data shows this pattern.
	(b) A case of the ratio increases monotonically as the baseline
		length increases.
	(c) A case of the ratio decreases almost monotonically as the
		baseline length increases.
	(d) A case of the ratio does not depend on the baseline length.
	}
\end{figure} 

To evaluate the effectiveness of the WVR phase correction on the long
baseline data, we first calculated the ratio of averaged rms phase
between the raw (not WVR corrected) and the WVR corrected data
(hereafter, we refer this as a phase improvement ratio) for each
baseline.
Fig~\ref{fig-ratio-bl} displays some example plots of the phase
improvement ratio as a function of baseline length.
Many (15 out of 25 data, or 60\%; Fig~\ref{fig-ratio-bl}a) of the
data show increment of the ratio as a function of baseline length up
to around several hundred meters to 1~km and then either being
constant or decrease at longer baselines.
On the other hand, small amount of data exhibit a constant increment
(5 out of 25, or 20\%; Fig~\ref{fig-ratio-bl}b), constant decrement
(2 out of 25, or 8\%; Fig~\ref{fig-ratio-bl}c), or even no trend
(3 out of 25, or 12\%; Fig~\ref{fig-ratio-bl}d).
This statistics tells that in most cases, the WVR phase correction
works effectively at some certain baseline length, but at longer
baseline length, this effectiveness being constant or decreases.
The baseline length of this effectiveness bending roughly matches
the turn-over baseline length in the structure functions (see next
section).
This suggests that the decrease of the effectiveness of the WVR phase
correction at the baseline lengths longer than the turn-over is due
to the change in the characteristics of phase fluctuation.
On the other hand, it can be due purely to the instrumental
characteristics of WVRs on the long baseline antennas, since the
long baseline data points rely only on a few antennas, and if one WVR
has lower sensitivity, then many long baseline data points will
appear to have low effectiveness of the WVR phase correction.

\begin{figure}
\begin{center}
\begin{tabular}{c}
\includegraphics[width=16cm]{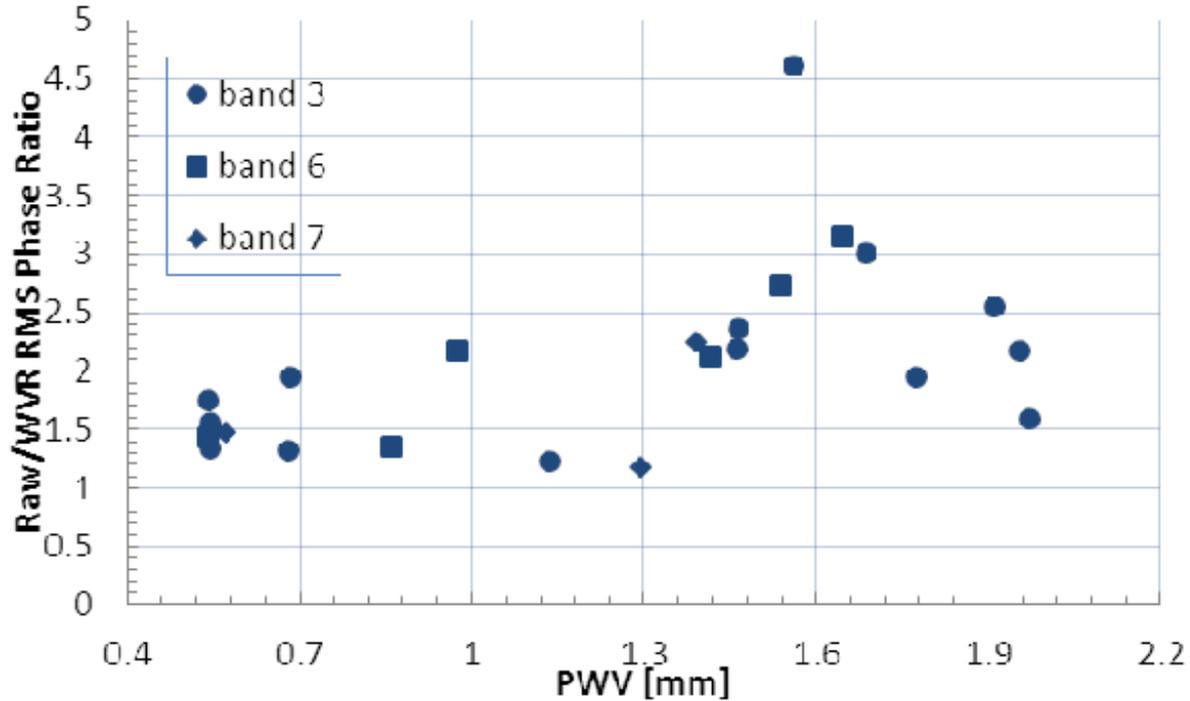}
\end{tabular}
\end{center}
\caption{
	\label{fig-ratio-pwv}
	Plot of the phase improvement ratio (the ratio of averaged rms
	phase between the raw and the WVR corrected data) as a function
	of precipitable water vapor (PWV).
	Circle, square, and diamond marks are the data from bands 3, 6,
	and 7, respectively.
	}
\end{figure} 

We then averaged over all the baselines in each data set.
This tells the overall degree of improvement of rms phase with
applying the WVR phase correction.
Fig.~\ref{fig-ratio-pwv} exhibits the averaged rms phase ratio between
the raw and the WVR-corrected data as a function of PWV.
There is a slight trend that higher PWV (PWV $>1$~mm) data improves
better than lower PWV (PWV $<1$ mm) data.
Indeed, the average ratio (i.e., improvement factor) for the higher
PWV data is 2.4, higher than that of the lower PWV data of 1.6.
This trend is consistent with that reported in the previous
papers\cite{mat12,nik13}, namely the WVR phase correction work well
under some amount of water vapor in the atmosphere, but will be
limited under drier (i.e., too less water vapor in the atmosphere)
conditions.
This figure also differentiates the frequency bands, but there is no
significant difference between the bands (although there are not much
data points for each band).

\section{Spatial Structure Function}
\label{sect-ssf}

\begin{figure}
\begin{center}
\begin{tabular}{c}
\includegraphics[width=16cm]{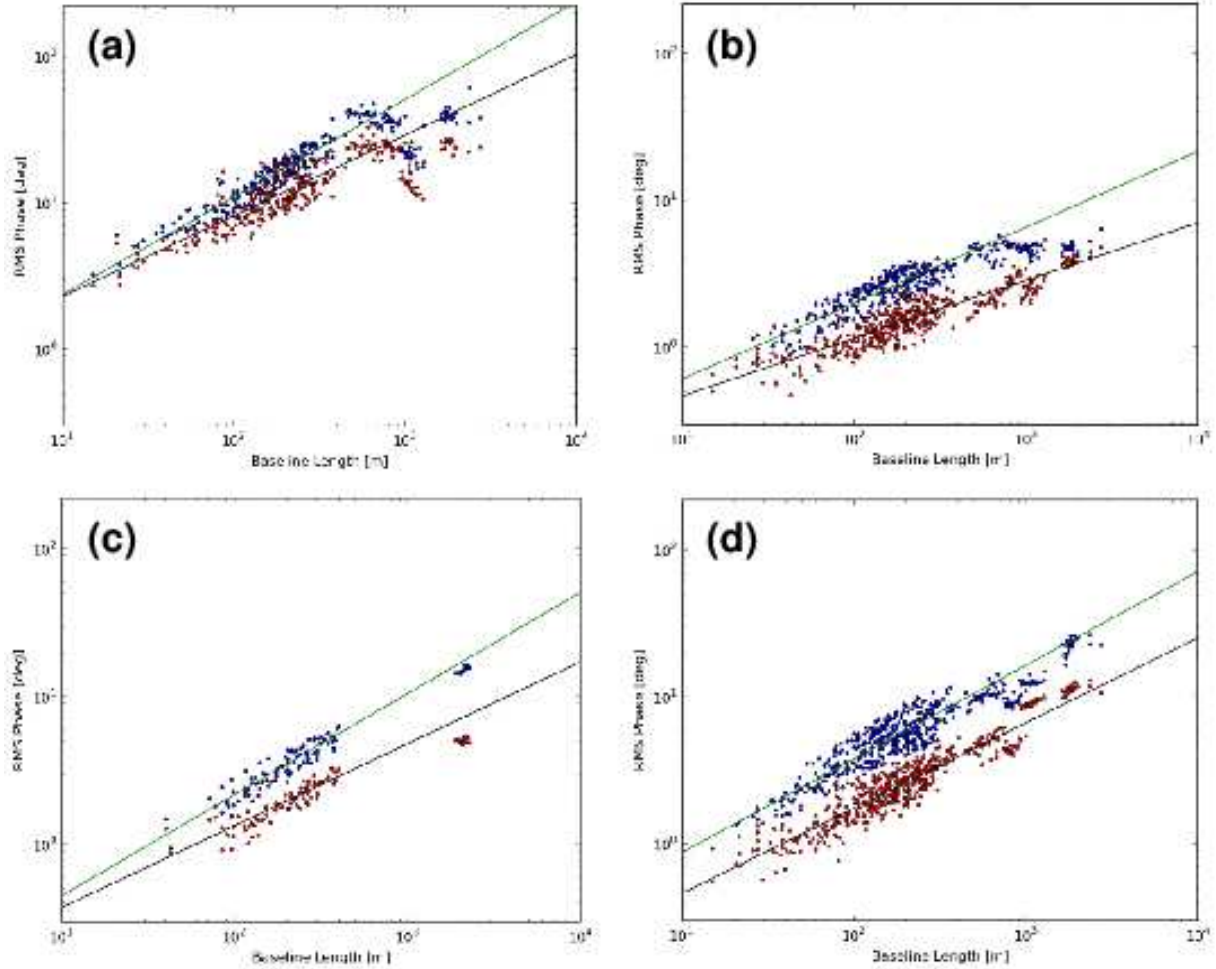}
\end{tabular}
\end{center}
\caption{
	\label{fig-ssf}
	Example plots of the spatial structure function (SSF) derived
	using the 3~km baseline data sets.
	Black and grey points are the raw and the WVR-corrected data
	points, respectively.
	Grey and black solid lines are the fitted slopes for the raw and
	the WVR-corrected data, respectively, using the data points at
	the baseline length shorter than 500~m.
	(a) A case that shows the turn-over in both the raw and
		WVR-corrected data.
	(b) A case that shows the turn-over only in the raw data.
	(c) A case that shows the turn-over only in the WVR-corrected
		data.
	(d) A case that does not show any turn-over in both the raw and
		the WVR-corrected data.
	}
\end{figure}

Fig.~\ref{fig-ssf} shows a few examples of SSF derived using the 3~km
baseline data sets.
Black and grey points are the raw and the WVR-corrected data points,
respectively.
For the first time, we have observed the turn-over of SSF in the
submillimeter-wave regime as in Fig.~\ref{fig-ssf}(a).
The turn-over is often observed around the baseline length of
several hundred meters to 1~km.
This suggests that for the baselines longer than this turn-over, the
phase fluctuation does not become larger than that measured around
the turn-over, and therefore assure the success of detecting fringes
at longer baseline, even for the longest baseline of ALMA of 15~km.

On the other hand, not all the data show the turn-over; some shows
either only in the raw data (i.e., turn-over does not exist in the
WVR-corrected data; Fig.~\ref{fig-ssf}b) or in the WVR-corrected
data (Fig.~\ref{fig-ssf}c), and some does not show in
both the raw and WVR-corrected data (Fig.~\ref{fig-ssf}d).
The data sets that show the turn-over in both the raw and the
WVR-corrected data are 3 out of 25 data sets (12\%), in only the raw
data 8 out of 25 data sets (32\%), in only the WVR-corrected data 1
out of 25 data sets (4\%), and that do not show the turn-over
are 13 out of 25 data sets (52\%).
This suggests that most of the WVR-corrected data do not exhibit the
turn-over.
However, the difficulty of judging the turn-over is that the number
of the long baseline data points is very limited due to the limited
number of antennas at the long baselines, so that the accuracy of the
location of the turn-over is not high.
Hereafter, we concentrate on the analysis of the slope of SSF.

\begin{figure}
\begin{center}
\begin{tabular}{c}
\includegraphics[width=16cm]{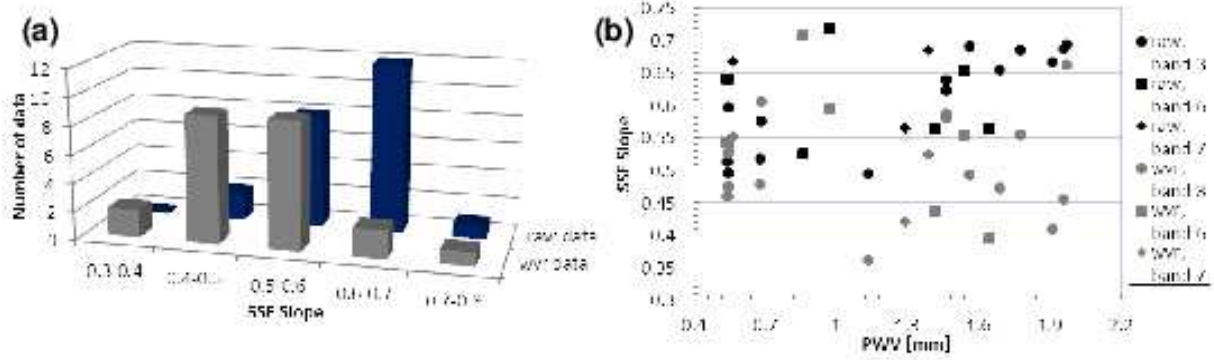}
\end{tabular}
\end{center}
\caption{
	\label{fig-ssf-slope}
	(a) Histogram of the number of the data sets as a function of
		fitted SSF slope for the raw (histogram behind) and
		WVR-corrected (histogram in front) data.
	(b) Fitted SSF slope as a function of PWV.
		Circle, square, and diamond marks are the data from bands 3,
		6, and 7, respectively.
		Black and grey color indicates the raw and WVR-corrected
		data, respectively.
	}
\end{figure} 

We fit the slope of SSF using the data points at the baseline length
shorter than 500~m.
The solid lines in Fig.~\ref{fig-ssf} are the fitted slopes.
Fittings have been done to all the data sets, and the histogram of
the number of the data sets as a function of slope are displayed in
Fig.~\ref{fig-ssf-slope}(a).
It is obvious that the peak of the histogram for the raw data are
larger (average = 0.61) than that of the WVR-corrected data (0.51),
namely the WVR-corrected data have shallower slope than that of the
raw data.
Note that the average (50\% quartile) slope of the temporal structure
function derived from the 11.2 GHz radio seeing monitor (RSM) data
used for the ALMA site testing\cite{but01} exhibits 0.63, agrees well
with our result for the raw data.
The RSM result is from the long term monitoring (taken between 1996
July and 1999 March), and the agreement of the average values for the
raw data therefore suggests that our data represent the typical phase
fluctuation condition (at least for the slope of the structure
function) at the ALMA site.

We then plot the slopes as a function of PWV
(Fig.~\ref{fig-ssf-slope}b).
For the data with the lower PWV of $<1$~mm, there is almost no
difference between the raw and the WVR-corrected data (average of
0.58 and 0.55, respectively), but for the data with the higher PWV of
$>1$~mm, the WVR-corrected data exhibit obviously shallower slopes of
0.49 than that for the raw data of 0.63.
We again differentiate the frequency bands in this figure, but there
is no significant difference between the bands; the aforementioned
trends are all true for all the bands.

Based on theoretical studies, steeper slope of 0.83 suggests the
three-dimensional Kolmogorov turbulence, and shallower slope of 0.33
suggests the two-dimensional Kolmogorov turbulence\cite{wri96,tho01}.
The slopes of our data for both raw and WVR-corrected ones are mostly
in between these values, although the WVR-corrected data have
shallower slopes than that of the raw data.
This suggests that the WVR phase correction makes the effect of
turbulence in phase fluctuation closer to two-dimensional turbulence.
On the other hand, if the phase fluctuation purely caused by the
water vapor in the atmosphere, the WVR phase correction will
eliminate all the phase fluctuation, and the slope should be flat
within the timescale of the WVR phase correction (which is 1 second).
All the obtained data exhibit, however, significant slopes,
suggesting other cause(s) for the atmospheric phase fluctuation.
The most likely cause of this is due to the dry component (O$_{2}$ or
N$_{2}$) in the atmosphere\cite{mat12,nik13}.
This, however, contradicts with the above discussion; the scale
height of the dry air ($\sim8$~km) is much larger than that of the
water vapor ($\sim2$~km), so that the turbulence in the dry air will
more likely to be three-dimensional, namely it seems natural to have
steeper slope.
But the obtained slopes tend to have shallower slopes, rather close
to the two-dimensional turbulence.
The alternative explanation may be the multiple layers of
two-dimensional Kolmogorov turbulence, which leads to the slopes to
be in the middle of two-dimensional and three-dimensional Kolmogorov
turbulence slopes.
This might be the most plausible solution, since the atmosphere
consists of multiple layers.
On the other hand, imperfect WVR phase correction due either to
instrumental causes or to incorrect assumptions in the phase
correction method cannot be ignored.
There is also a possibility of phase fluctuation caused by
instruments, but it is usually difficult to produce baseline-base
phase fluctuation\cite{mat12}, so it is less likely.

\section{Coherence Time}
\label{sect-cohT}

\begin{figure}
\begin{center}
\begin{tabular}{c}
\includegraphics[width=16cm]{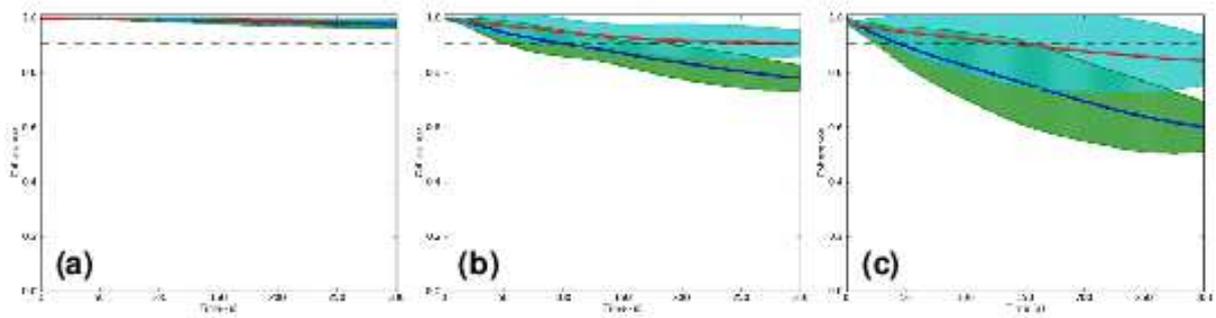}
\end{tabular}
\end{center}
\caption{
	\label{fig-cohTime}
	Example plots of the coherence as a function of time in second.
	Here we define the coherence time as the coherence degrade for
	10\% (i.e., coherence of 0.9) with integrating the data, namely
	due to phase decorrelation.
	Coherence of 0.9 is displayed with a dashed line in each plot.
	Black and grey points are the raw and the WVR-corrected data,
	respectively.
	Error bars are also plotted in each data point.
	(a) An example plot of band 3 data.
		In this plot, both the raw and WVR-corrected data do not
		decorrelate to the coherence of 0.9.
	(b) An example plot of band 6 data.
	(c) An example plot of band 7 data.
	In both bands 6 and 7 data, WVR-corrected data have longer
	coherence time than the raw data.
	}
\end{figure}

Since some of the data show the turn-over in the slope of SSF, it is
worth to calculate the coherence time at the longest baseline length
of 3~km, because it assures that there is no increment of the
coherence time even for longer baselines (even for the very long
baseline interferometry [VLBI] observations, which is considered as
one of the near future plans for ALMA).
Here we assume that the coherence time corresponds to the amplitude
decrement of 10\% from the original amplitude due to the phase
decorrelation.
We only use the 3~km baseline length data for this calculation.
We also calculate the coherence time for all the frequency bands
using the original data, assuming that the phase fluctuation
linearly increases with the frequency increase (i.e., non-dispersive
assumption).
This assumption is valid, since the WVR phase correction works at all
the frequency bands in our data, which also assumes the
non-dispersive characteristics of the phase at the observation
frequencies.
Due to the limited observation time (and therefore the statistics for
each time length), we limit the calculation of the coherence time up
to $\sim400$ seconds.

\begin{figure}
\begin{center}
\begin{tabular}{c}
\includegraphics[width=16cm]{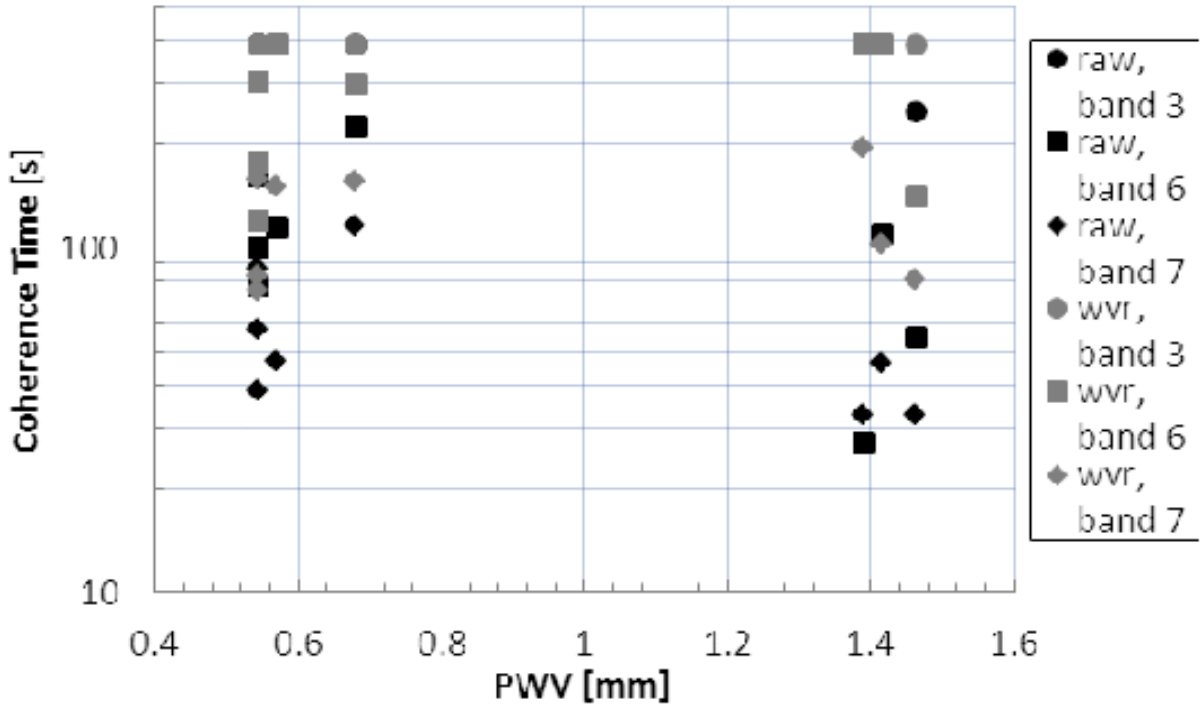}
\end{tabular}
\end{center}
\caption{
	\label{fig-cohTime-pwv}
	Coherence time as a function of PWV.
	Circle, square, and diamond marks are the data from bands 3, 6,
	and 7, respectively.
	Black and grey color indicates the raw and WVR-corrected data,
	respectively.
	Data points around the coherence time of 400 seconds are the data
	that do not degrade to the coherence of 0.9, namely the coherence
	time of $\sim400$ seconds are the lower limit for these data.
	This limitation is caused by the limited measurement time length.
	}
\end{figure}

An example plot of coherence as a function of time for each band is
displayed in Fig.~\ref{fig-cohTime} and the coherence time as a
function PWV is displayed in Fig.~\ref{fig-cohTime-pwv}.
Most (7 out of 8) of the raw data and all of the WVR-corrected data
of band 3 have coherence time longer than $\sim400$ seconds, namely
it is possible to have a switching time between a target source and
a calibrator of at least 6.5 minutes for the most of the data (as far
as the PWV of less than 2~mm; no data so far for PWV larger than
2~mm) without significant phase decorrelation for band 3.
This is also true for VLBI observations; it is possible to integrate
at least 6.5 minutes for the most of the band 3 data at the ALMA
site.

For bands 6 and 7, the average coherence time is 108 and 57 seconds,
respectively, for the raw data and $>273$ and 125 seconds for the WVR
corrected data.
This tells that for these higher frequency observations, especially
for band 7 (or higher frequency bands), it is better to observe with
fast switching calibration technique to keep coherence loss less than
10\%.
This results do not reject to observe slow switching observation, but
the data taken with the slow switching will have some degree (more
than 10\%) of coherence loss.

It is obvious that the WVR phase correction extends the coherence
time significantly (more than a factor of 2) from the non-WVR
corrected data.
This also suggests that the WVR phase correction helps a lot for VLBI
observations.
But still it is better to operate with fast switching calibration,
especially for higher frequency bands (higher than band 7) to have
low coherence loss data.


\acknowledgments     

We would like to thank the Joint ALMA Observatory (JAO) staff for
supporting our measurements.
S.M.\ is supported by the National Science Council (NSC) of Taiwan,
NSC 100-2112-M-001-006-MY3.



\begin{thebibliography}{1}

\bibitem{hil10}
R.~E.~Hills,  R.~J.~Kurz, and A.~B.~Peck,
  ``ALMA: status report on construction and early results from
    commissioning,''
  {\em Proc.~SPIE} {\bf 7733}, pp.~773317-773317-10, 2010.

\bibitem{tho01}
A.~R.~Thompson, J.~M.~Moran, and G.~W.~Swenson,~Jr.,
  {\em Interferometry and Synthesis in Radio Astronomy},
  Wiley-Interscience, New York, 2001 (second edition).




\bibitem{nik13}
B.~Nikolic, R.~C.~Bolton, S.~F.~Graves, R.~E.~Hills, and J.~S.~Richer,
  ``Phase correction for ALMA with 183 GHz water vapour radiometers,''
  {\em Astron.\ \& Astrophys.} {\bf 552}, A104, pp.~11, 2013.

\bibitem{mat12}
S.~Matsushita, K.-I.~Morita, D.~Barkats, R.~E.~Hills, E.~B.~Fomalont,
  and B.~Nikolic,
  ``ALMA temporal phase stability and the effectiveness of water vapor
    radiometer,''
  {\em Proc.~SPIE} {\bf 8444}, pp.~84443E-84443E-7, 2012.

\bibitem{mcm07}
J.~P.~McMullin, B.~Waters, D.~Schiebel, W.~Young, and K.~Golap,
  ``CASA Architecture and Applications,''
  {\em ASP Conf.~Ser.} {\bf 376}, pp.~127-130, 2007.

\bibitem{tat61}
V.~I.~Tatarskii,
  {\em Wave Propagation in a Turbulent Medium}, Dover, New York, 1961.

\bibitem{hol95}
M.~A.~Holdaway, S.~J.~E.~Radford, F.~N.~Owen, and S.~M.~Foster,
  ``Data Processing for Site Test Interferometers,''
  {\em ALMA Memo} {\bf 129}, 1995.

\bibitem{sra06}
R.~Sramek, and C.~Haupt,
  ``ALMA System Technical Requirements for 12m array,''
  {\em ALMA-80.04.00.00-005-B-SPE}, 2006.

\bibitem{but01}
B.~J.~Butler, S.~J.~E.~Radford, S.~Sakamoto, and K.~Kohno
  ``Atmospheric Phase Stability at Chajnantor and Pampa la Bola,''
  {\em ALMA Memo} {\bf 365}, 2001.

\bibitem{wri96}
M.~C.~H.~Wright,
  ``Atmospheric Phase Noise and Aperture Synthesis Imaging at
	Millimeter Wavelengths,''
  {\em Publ.\ Astron.\ Soc.\ Pacific} {\bf 108}, pp.~520-534, 1996.

\end{thebibliography}
\end{document}